\documentstyle[epsfig]{IOPConf}
\begin{document}
\frenchspacing
\newcommand{\bm}{\bibitem}
\newcommand{\ud}{\bf}
\renewcommand{\thefootnote}{\fnsymbol{footnote}}

\title{Structure of $^8$B and astrophysical $S_{17}$ factor}
\longauthor{S.K. Dhiman\dag ~ and R. Shyam\ddag}
        {S.K. Dhiman and R. Shyam}
\address{\dag Department of Physics, Himachal Pradesh University, Shimla 171 005, India\\
\ddag Saha Institute of Nuclear Physics, Kolkata - 700 064, India.}

\beginabstract
Nuclear structure data are of crucial importance in order to address
important astrophysical problems such as the origin of chemical elements,
the inner working of our Sun, and the evolution of stars. We demonstrate 
this by investigating the ground state structure of $^8$B and $^7$Be 
nuclei within the Skyrme Hartree-Fock framework and by calculating the   
overlap integral of $^8$B and $^7$Be wave functions. The latter is used
to calculate the astrophysical S factor ($S_{17}$) for the solar 
fusion reaction $^{7}$Be($p, \gamma)^{8}$B.
\endabstract
\section{Introduction}
Nuclear astrophysics research addresses some of the most fundamental
questions in nature {\it e.g.,} the origin of the elements that make up
our bodies and our world, and formation and evolution of the sun, the stars
and the galaxies. There is an intimate connection between
nuclear physics inputs and studies of these fascinating astrophysical
phenomena~\cite{cag05}. A diverse set of nuclear data is required to model the 
composition changes and energy release in astrophysical environments
ranging from Big Bang to inner working of our own Sun to exploding stars.
Theoretical studies as well as measurements of microscopic nuclear physics
processes provide the foundation for models to understand various 
astrophysics phenomena. These models are now challenged by incredibly
detailed observations from ground based and space based ({\it e.g.,} 
CHANDRA X-ray Observatory, Hubble Space Telescope) observational 
devices that give us an unprecedented view of the Cosmos

We shall demonstrate the role of the nuclear structure physics in 
nuclear astrophysics by considering the case of the Solar fusion reaction
$^{7}$Be(p,$\gamma)^{8}$B, which is of great importance to the solar
neutrino issue and to other related astrophysical studies. $^8$B
is the source of the high energy neutrinos from the Sun that are detected
in the SNO, Kamiokande and Homestake experiments \cite{bac89,dav94,fuk01}. 
It is, therefore, a crucial requirement to determine as accurately as
possible the cross section of this reaction at relative energies
corresponding to solar temperatures (about 20 keV).  In this energy region,
the cross-section $\sigma_{p \gamma}(E_{cm})$ [which is usually expressed
in terms of the astrophysical $S_{17}(E_{cm})$ factor] of the
$^7$Be(p,$\gamma)^8$B reaction is directly proportional to the
high energy solar neutrino flux. A better knowledge of $S_{17}$ is,
therefore, important to improve the precision of the theoretical prediction
of $^8$B neutrino flux from present and future solar neutrino experiments.

$S_{17}(0)$] is determined either by direct measurements \cite{ham01,jun03}
or by indirect methods such as Coulomb dissociation \cite{mot94} 
and transfer reactions~\cite{gag01,das04}. Efforts have also
been made to calculate the cross section of this reaction within the
framework of the shell model and the cluster model \cite{cso00,bro96}.
The key point of these calculations is the determination of the 
wave functions of $^8$B states within the given structure theory.
 
We have investigated the structure of $^8$B
in the framework of the Skyrme Hartree-Fock (SkHF) model which has
been used successfully to describe the ground-state properties of
both stable~\cite{que78,dob94} as well as exotic nuclei 
\cite{miz00}. The SkHF method with density-dependent
pairing correlation and SLy4 interaction parameters has been successful
in reproducing the binding energies and rms radii \cite{miz00} in the
light neutron halo nuclei $^6$He, $^8$He, $^{11}$Li and $^{14}$Be.
We solve spherically symmetric Hartree-Fock (HF) equations with SLy4
\cite{cha97} Skyrme interaction which has been
constructed by fitting to the experimental data on
radii and binding energies of symmetric and neutron-rich nuclei.
Pairing correlations among nucleons have been
treated within the BCS pairing method. We have, however, renormalized
the parameter of the spin-orbit term of the SLy4 interaction so as to
reproduce the experimental binding energy of the last proton in the $^{8}$B
nucleus \cite{sha93}. A check on our interaction parameters was
made  by calculating binding energies and rms radii of $^7$Be, $^7$B,
$^8$Li and $^9$C nuclei with the same set where a 
good agreement is obtained with corresponding experimental data.
The overlap integral of the HF wave functions for $^7$Be and $^8$B ground
states has been  used to calculate the astrophysical $S_{17}$ factor.

\section{Results and discussions}

\subsection{Structure calculations}
 
The values of various parameters of the $SLy4$ Skyrme
effective interaction as used in our calculations are given in \cite{sha93}. 
The rms radii for matter ($r_m$), neutron ($r_n$) and proton ($r_p$)
distributions are presented in Table I for five light nuclei. Also
shown in this table are the matter, neutron, and proton
radii (under the column "Expt") extracted by methods in which measured
reaction (or interaction) cross sections are fitted by theoretical
models having them as input parameters. The quantities listed under
"Theory" column are the results of our calculations. Here $r_m$ is
obtained by  summing the average of proton and neutron radii in every
orbit weighted with occupation probabilities.  We see that
for all the isotopes the calculated $r_m$ is in good agreement with the
corresponding values listed under the "Expt" column. 
\begin{table}[here]
\caption{ Rms mass ($r_m$), proton ($r_p$) and neutron ($r_n$) radii for
various nuclei. Theoretical results obtained with modified SLy4 Skyrme
force. Under the column "Expt", are the values of corresponding radii
extracted by fitting the reaction or interaction cross sections by
different theoretical methods as discussed in the text. We have defined
$r_i = <r_i^2>^{1/2}$}
\vspace{0.3cm}
\begin{tabular}{cccccccc}
\hline
\vspace{0.1cm}
Nucl. & \multicolumn{6}{c}{rms radii(fm)}\\ 
    & \multicolumn{3}{c}{"Expt"} & \multicolumn{3}{c}{Theory}\\
\hline
 & $r_{m}$ & $r_{p}$ & $r_{n}$ & $r_{m}$
 & $r_{p}$ & $r_{n}$\\ 
\hline
\\
 $^{7}Be$ &$2.33\pm0.02$ & - & - &2.49 &2.63 &2.29 \\
 $^{7}B$  &-  &- &- &2.86 &3.18 &1.84  \\
 $^{8}Li$ &$2.37\pm0.02$ &$2.26\pm0.02$ &$2.44\pm0.02$ &2.54 &2.29 &2.67\\
 $^{8}B$  &$2.55\pm0.08$ &$2.76\pm0.08$ &$2.16\pm0.08$ &2.57 &2.73 &2.27\\
   &$2.43\pm0.03$ &$2.49\pm0.03$ &$2.33\pm0.03$ & & & \\
 $^{9}C$  &$2.42\pm0.03$ & & &2.59 &2.77 &2.20 \\
\hline
\end{tabular}
\end{table}

%In Fig.~1, we show density distributions [$\rho (r)$] of proton,
%the valence particle and the core. $\rho (r)$  
%has been obtained by folding the HF results for proton and neutron
%densities with the intrinsic charge density distribution of nucleons
%in the Fourier-space by transforming the densities to form-factors.
%We note that the proton density distribution closely follows that of
%the valence particle. However, they differ from that of the core at 
%larger distances. This indicates that the valence nucleon density 
%distribution gets decoupled from that of the core at distances
%larger than 3 fm. This is reminiscent of the situation in the neutron
%halo nuclei. This observation
%supports the existence of a proton halo structure in $^8$B.
%
%\begin{figure}[htb]
%\begin{center}
%\epsfig{file=rho_in.eps,height=7.0cm }
%\end{center}
%\caption[C1]{\captionfont {Density distribution $\rho(r)$ for the proton,
%the valence nucleon  and the core
%in the $^{8}B$ nucleus calculated with SkHF method.}} 
%\label{fig:figa}
%\end{figure}

\subsection{ Valence proton radius in $^{8}B$ and Astrophysical
$S_{17}$ factor }

We define a overlap function of the bound state wave functions of two
nuclei $B$ and $A$, where $B = A + p$ ($p$ represents a proton) as  
\begin{eqnarray}
I_A^B ({\bf r}) & = & \int d\xi \Psi_{A I_A M_A}^*(\xi) \Psi_{B I_B M_B}
({\bf r},\sigma_p,\xi),
\end{eqnarray}
where $I_A$ and $I_B$ are the total spins of nuclei A and B, respectively,
and ${\bf r}$ is the position of the proton with respect to the 
c.m. of nucleus A. $\sigma_p$ is the spin variable of the proton, and $\xi$
stands for the remaining set of internal variables which also include
isospins. In this expression nuclear wave functions $\Psi_{A}$ and $\Psi_{B}$
are supposed to be properly translational invariant. The Hartree-Fock wave
functions calculated in the previous section may not be so despite the fact
that a c.m. correction factor  has been incorporated in the energy
functional. Corrections for the spurious c.m. motion should, therefore, be
applied to the HF wave functions before using them in Eq.~(1). However,
effects of such corrections on the one-particle overlap function
calculated within the shell model have been found~\cite{pink76,rsh83}
to be of the order of only about 2-5 $\%$. It is quite likely that the
situation will be no different for the HF case. 

We, now, present results for the astrophysical $S_{17}$ factor
calculated using HF overlap functions. 
In the region outside the core where range of the nuclear interaction
becomes negligible, the radial overlap wave function
of the bound state can be written as  
\begin{eqnarray}
R_{A\ell j}^B(r) & \simeq & \bar{c_{lj}}W_{\eta ,l+1/2}(2kr)/r,
\end{eqnarray}
where $W$ is the Whittaker function, $k$ the wave number corresponding
to the single proton separation energy and $\eta $  the Sommerfield
parameter for the bound state. In Eq.~(2), 
$\bar{c_{lj}}$ is the asymptotic normalization constant, required
to normalize the radial overlap wave function 
to the Whittaker function in the asymptotic region. 
The $S_{17}$ factor is related to the proton capture
cross section as  
\begin{equation}
S_{17}(E)=\sigma (E)E\,e^{(2\pi \eta (E))}.
\end{equation}
At the zero energy the $S_{17}$ factor depends only on $\bar{c_{lj}}$
and one can write \cite{cso00,bro96} 
\begin{equation}
S_{17}(0)=\kappa\sum_{j}\bar{c}_{1 j}^{2},
\end{equation}
In Ref.~\cite{cso00}, $\kappa$ (= 37.8) has been obtained by using 
a microscopic cluster model for the scattering states, while a value of 36.5 
has been reported for this quantity in Ref.~\cite{bro96} using a hard sphere
scattering state model. However, once the relevant integration distances
are sufficiently enhanced in~\cite{bro96} the value of $\kappa$ there
comes out to be 37.2, which is in good agreement with that of the
microscopic model.

In our calculations, we have used our HF overlap function directly
in the calculation of the amplitudes for the direct capture reaction.
The amplitude for the radiative capture reaction can be written as
$ b+c \to a + \gamma$ 
which can be written as
\begin{eqnarray}
M_{fi} & = & \langle\phi_a(\xi_b,\xi_c,r)|{\hat O}(r)
|\phi_b(\xi_b)\phi_c(\xi_c) \psi_{k_i}^{(+)}({\bf r})\rangle,
\end{eqnarray}
where $\phi$, $\xi$ and ${\bf r}$ are the wave function, and the internal
coordinate of the bound particle and the relative coordinate between
$b$ and $c$, respectively. ${\hat O}(r)$ is the electromagnetic
operator. $ \psi_{k_i}^{(+)}({\bf r})$ is the distorted wave in
the initial $b + c$ channel. The overlap of initial
and final bound state wave functions is defined by
\begin{eqnarray}
I_{bc}^a({\bf r}) & = &\langle\phi_a(\xi_b,\xi_c,r)|\phi_b(\xi_b)\phi_c(\xi_c)
\rangle,\nonumber \\
 & = & \sum_{\ell j \mu} <I_b M_b j m | I_a M_a>
<\ell m-\mu s \mu | j m> i^\ell R_{b\ell j}^a (r) Y_{\ell m-\mu}({\hat r})
\nonumber \\
& \times & \chi_{s\mu},
\end{eqnarray}
It may be noted that at low relative energies the distortion effects are
goverened solely by Coulomb interactions. Substituting Eq.~(6) into
Eq.~(5), expanding $ \psi_{k_i}^{(+)}({\bf r})$ in terms of the partial
waves and carrying out the angular momentum algebra, the amplitude
$M_{fi}$ can be written in terms of the wave function $R_{b\ell j}^a (r)$
and the radial coulomb wave functions for the initial channel. Therefore,
the overlap functions $R_{b\ell j}^a (r)$ calculated in the HF theory
can be used to calculate the capture cross sections at low 
relative energies without any uncertainty of other input quantity. 
It may further be noted that using Eq.~(2), we can also 
calculate the asymptotic normalization constants ${\bar c}_{\ell j}$. 

In Table II, we show the results for the asymptotic normalization 
constants for the $p_{1/2}$ and $p_{3/2}$ proton orbits and the 
the astrophysical $S$-factor obtained by this method.  
Our ${\bar c}_{3/2}$ agrees almost perfectly with the value of this
quantity determined experimentaly in Ref.~\cite{tra03}. However,
our ${\bar c}_{1/2}$ is larger than that reported by these
authors by a factor of about 1.5. They have determined 
these quantities (by invoking mirror symmetry) from the ANCs of
the $^8$Li $\to$ $^7$Li $+ n$ system extracted from the measurement of
the neutron transfer reaction $^{13}$C($^7$Li,$^8$Li)$^{12}$C.
This difference in the calculated and "measured" values of
${\bar c}_{1/2}$ is significant as a larger ${\bar c}_{1/2}$
leads to a bigger $S_{17}$ as compared to that reported in
Ref.~\cite{tra03}. There may be a need to relook in the theoretical
analysis of the data reported in Ref.~\cite{tra03}. In the DWBA
calculations of the $^{13}$C($^7$Li,$^8$Li)$^{12}$C reaction, same
set of potentials have been used for $^7$Li + C and $^8$Li + C channels.
However, $^8$Li is an unstable system; thus optical potential in the final 
channel could be different from that of the incoming ($^7$Li) channel.
It is also not very clear from Ref.~\cite{tra03} if the spin-orbit
terms have been used in the optical potentials and bound state 
potentials. This could affect the $p_{1/2}$ and $p_{3/2}$ components
differently. More work is clearly needed to clarify this issue.  

\begin{table}[here]
\begin{center}
\caption{SkHF results for    
asymptotic normalization coefficients
($\bar{c}_{\ell j}$), and astrophysical $S-$ factors $S^A_{17}(0)$
and $S^B_{17}(0)$ and $S^C_{17}(0)$.
$S_{17}^{A}$ corresponds to results obtained by using the HF
overlap function directly to a capture code
while $S_{17}^{B}$ and $S_{17}^C$ to those
obtained by using Eq.~(4) with $\kappa$ = 36.5 and 37.8, respectively.}
\vspace{0.3cm}
\begin{tabular}{cc}
\hline
\\
Observable & Our Result \\
\hline
\\
$\bar{c}_{1 3/2}$ & 0.64\\
$\bar{c}_{1 1/2}$ & 0.34  \\
$S^{A}_{17}(0) (eV b)$ & 22.0 \\
$S^{B}_{17}(0) (eV b)$ & 19.5 \\
$S^{C}_{17}(0) (eV b)$ & 20.2 \\
\\
\hline 
\end{tabular}
\end{center}
\end{table}

The recent  $^7$Be($p,\gamma$) measurements yield
values of $S_{17}(0)$  which  are  clustered  around  18.5~eV~b
\cite{ham01} and 22.0~eV~b~\cite{jun03}. Our $S_{17}(0)$ (22.0 eV barn)
is closer to two latest direct capture measurement results 
which claim good accuracy. However, this is larger than the most recent
result reported by the Texas A \& M group from the ANC method~~\cite{tra03}.
This is also slightly larger than the values obtained from the
Coulomb dissociation (CD) method~\cite{esb05}, even thought it has been 
argued~\cite{gai05} that there is no significant difference between the CD 
and diect capture values of $S_{17}(0)$. 
\section{Summary and conclusions}

In summary, in this paper we studied the structure of $^8$B,
and $^7$Be nuclei within the Skyrme Hartree-Fock (SkHF) 
framework. We calculated binding energies, various densities distribution
and rms radii for these nuclei. Using the same set of the force parameters,
we obtain good agreements with experimental values of binding energies and
rms matter radii for all these nuclei. We have calculated the overlap
function $<^7$Be$\mid ^8$B$>$ from the SkHF wave functions which has been
employed to extract the asymptotic normalization coefficients for the 
$^8$B $\rightarrow ^7$Be $+p$ system.
We obtain an astrophysical S-factor of 22.0 eV b which lies within the 
adopted limits ($19.1^{+4.0}_{-1.0}$ eV b) of the near zero
energy astrophysical S-factor. Our work clearly shows that proper nuclear
structure input is vital in order to understand some very important issues
in nuclear astrophysics. 
\section{ Acknowledgment}

This work is supported by the Department of Atomic Energy,
(BRNS), (BARC), Mumbai, under contract no. 2001/37/14/BRNS/699.

%\begin{thebibliography}{99}
\references
                                                                                
\list
 {[\arabic{enumi}]}{\settowidth\labelwidth{[99]}\leftmargin\labelwidth
 \advance\leftmargin\labelsep
 \usecounter{enumi}}
 \def\newblock{\hskip .11em plus .33em minus .07em}
 \sloppy\clubpenalty4000\widowpenalty4000
 \sfcode`\.=1000\relax
 \let\endthebibliography=\endlist
                                                                                
\itemsep=-1pt

\bibitem{cag05} Jac Caggiano, these proceedings.

\bibitem{bac89}
J.N. Bahcall {\it Neutrino Astrophysics} (Cambridge University Press, 1989),
 J.N. Bahcall and M. Pinsonneault, Rev. Mod. Phys. {\bf 64} (1992) 885.

\bibitem{dav94}
R. Davis, Prog. Part. Nucl. Phys. {\bf 32} (1994) 13,
Eric G. Adelberger et al., Rev. Mod. Phys. {\bf 70} (1998) 1265.

\bibitem{fuk01}
S. Fukuda et al, Phys. Rev. Lett. {\bf 86} (2001) 5651,
Q.R. Ahmad et al., {\bf 87} (2001) 071031, {\bf 89} (2002) 011301.

\bibitem{ham01}
F. Hammache et al., Phys. Rev. Lett {\bf 86} (2001) 3985.

\bibitem{jun03}
A.R. Junghans et al., Phys. Rev. Lett. {\bf 88} (2002) 041101;
L. T. Baby {\it et al}., Phys. Rev. Lett. {\bf 90}, (2003) 022501.

\bibitem{mot94}
I. Motobayashi, Nucl. Phys. A{\bf 693} (2001) 258c;
B. Davids, et al., Phys. Rev. C{\bf 63} (2001) 065806.

\bibitem{gag01}
G.A. Gagliardi et al., Nucl. Phys. {\bf A682} (2001) 369.

\bibitem{das04}
J. J. Das et al., Nucl. Phys. {\bf A746} (2004) 561c.

\bibitem{cso00}
A. Csoto, Phys. Rev. C{\bf 61} (2000) 037601;
B.K. Jennings, S. Karataglidis, and T.D. Shoppa, Phys. Rev. C {\bf 58}
(1998) 3711.

\bibitem{bro96}
B.A. Brown, A. Costo, and R. Sherr, Nucl. Phys. A{\bf 597} (1996) 66.

\bibitem{que78}
F. Quentin and H. Flocard, Annu. Rev. Nucl. Part. Sci. {\bf 28} (1978) 523.

\bibitem{dob94}
J. Dobaczewski, I. Hamamoto, W. Nazarewicz and J.A. Sheikh, Phys. Rev. Lett. 
{\bf 72} (1994) 981.

\bibitem{miz00}
S. Mizutori, J. Dobaczewski, G. A. Lalazissis, W. Nazarewicz
and P. G. Reinhard, Phys. Rev. C{\bf 61} (2000) 044326.

\bibitem{cha97}
E. Chabanat, P. Bonche, P. Haensel, J. Meyey and R. Schaeffer,
 Nucl. Phys. A.{\bf 627}, 710(1997).

\bibitem{sha93}
S.S. Chandel, S.K. Dhiman, and R. Shyam, Phys. Rev. C{\bf 68} (2003) 054320.

\bibitem{pink76}
W. T. Pinkston, Nucl. Phys. {\bf A269} (1976) 281; W. T. Pinkston and
P. J. Iano, Nucl. Phys. {\bf A330} (1979) 91.

\bibitem{rsh83}
R. Shyam and M. A. Nagarajan, J. Phys. G: Nucl. Part. Phys. {\bf 9}
(1983) 901.

\bibitem{tra03}
L. Trache et. al., Phys. Rev. C {\bf 67} (2003) 062801(R).

\bibitem{esb05}
H. Esbensen, G.F. Bertsch, and K. A. Snover, Phys. Rev. Lett. {\bf 94}
(2005) 042502.

\bibitem{gai05}
M. Gai, nucl-ex/0502020.
%\end{thebibliography}
\endlist
\end{document}